\documentclass{jaa}
\usepackage{natbib}
\usepackage{graphicx}
\usepackage{amsmath}
\usepackage[caption=false]{subfig}
\usepackage{hyperref}

\begin{document}\sloppy

\title{UV photometry of spotted stars in the horizontal branch of the globular cluster NGC\,2808 using \textit{AstroSat}}


\author{Deepthi S. Prabhu\textsuperscript{1,2*}, Annapurni Subramaniam\textsuperscript{1} and Snehalata Sahu\textsuperscript{1}}
\affilOne{\textsuperscript{1}Indian Institute of Astrophysics, Bangalore.\\}
\affilTwo{\textsuperscript{2}Pondicherry University, R.V. Nagar, Kalapet, 605014, Puducherry, India.}


\twocolumn[{

\maketitle

\corres{deepthi.prabhu@iiap.res.in}

\msinfo{}{}

\begin{abstract}
A recent study of hot (20,000 to 30,000\,K) extreme horizontal branch (EHB) stars in globular clusters (GCs) has led to the discovery of their variability. It is suggested that this variability is driven by the projected rotation of magnetic spots on the stellar surfaces and is expected to have higher amplitudes at shorter wavelengths. Here, we present the analysis of such hot stars in the massive GC NGC\,2808 using the Ultraviolet Imaging Telescope (UVIT), aboard \textit{AstroSat}. We use the UVIT data in combination with the {\it Hubble Space Telescope} UV Globular Cluster Survey (HUGS) data for the central region (within $\sim$ $2.7'$ $\times$ $2.7'$) and ground-based optical photometry for the outer parts of the cluster. We generate the far-UV (FUV) - optical colour-magnitude diagrams (CMDs) and in these we find a population of hot EHB stars fainter than the zero-age horizontal branch (ZAHB) model. A comparison of our FUV magnitudes of the already reported variable EHB stars (vEHBs) shows that the longest period vEHBs are the faintest, along with a tentative correlation between rotation period and UV magnitude of spotted stars. In order to firmly establish any correlation, further study is essential.
\end{abstract}


}]


\doinum{12.3456/s78910-011-012-3}
\artcitid{\#\#\#\#}
\volnum{000}
\year{0000}
\pgrange{1--}
\setcounter{page}{1}
\lp{1}

\section{Introduction}
Horizontal branch (HB) stars are low-mass Helium (He) core burning stars consisting of a hydrogen-rich envelope. Globular clusters (GCs) are the best test beds to explore low-mass stars in various stages of evolution, including the HB stars. The morphology of the HB in GCs shows many peculiarities, one of which is the well-known ``second parameter'' problem \citep{SandageWallerstein1960,SandageWildey1967,vandenbergh1967}. This refers to an observation that the colour distribution of HB stars is affected by parameters (eg., age, He abundance) other than metallicity \citep{Catelan2009}. 

At a given metallicity, the colour distribution of the HB is a function of the envelope masses of stars which in turn depends on the mass-loss on the red giant branch (RGB) phase \citep{IbenRood1970, Rood1973}. The less massive the envelope is, the hotter the resulting star is. The HB is thus made up of various sub-populations having similar core masses ($\sim$ 0.5 $M_{\odot}$) with differences in envelope masses. These include red HB (RHB), blue HB (BHB), extreme HB (EHB) and blue hook (BHk) stars or late hot flashers \citep{Dcruz1996,Dcruz2000,Sweigart1997}, in the increasing order of effective temperatures ($T_{eff}$). The hot HB stars along with post-HB (pHB) stars, white dwarfs (WDs) and the exotic blue straggler stars (BSSs) emit copious amount of flux in the ultraviolet (UV) wavebands. 

A recent study by \citet{Momany2020} utilizing near-UV (NUV) and optical data, led to the discovery of variable stars among the EHB population (vEHB), in three Galactic GCs, namely, NGC\,2808, NGC\,6752 and NGC\,5139 ($\omega$ Cen.). The EHB stars have extremely thin radiative envelopes ($\lesssim$ 0.02 $M_{\odot}$) and $T_{eff}$ ranging from $\sim$ 20,000 to 30,000\,K. Their counterparts in the Galactic field are sub-dwarf B-type (sdB) stars. Due to their thin envelopes, the EHB stars do not ascend the asymptotic giant branch (AGB) phase after core He exhaustion. Instead, they briefly brighten up as AGB-manqu\'e stars and then evolve directly towards the WD phase. 

\citet{Momany2020} reported two types of EHB variability, periodic (periodicity of $\sim$ 2 to 50 days with $\Delta$ $U_{Johnson}$ $\approx$ 0.04 to 0.22 mag) and aperiodic. This EHB variability was observed to be tightly connected around the photometric jump at $T_{eff}$ $\sim$ 22,500\,K (the Momany jump, \citet{Momany2002}), a feature observed in all GCs \citep{Momany2004, Brown2016}. Through appropriate analyses and arguments, binary evolution or pulsation were discarded as reasons for the variability. The origin of both modes of EHB variability was attributed to the $\alpha^{2}$ Canum Venaticorum ($\alpha^{2}$\,CVn) phenomenon. In this scenario, chemical inhomogeneities on the stellar surface cause spatial variations in opacity resulting in spots which are stabilized for a long time (several decades) by an underlying magnetic field. The projected rotation of these magnetic spots causes photometric/spectroscopic variability. The uneven surface chemical distribution in EHB stars is caused due to atomic diffusion processes. The magnetic fields required to sustain the spots were argued to be resulting from the second ionization He convective zone (HeIICZ), lying just beneath the radiative envelope. The aperiodic variability was explained as a consequence of the turbulent nature of the HeIICZ. Their work established that magnetism plays a key role in the formation and evolution of EHB stars and their Galactic field counterparts.

The magnetic spots caused by the $\alpha^{2}$\,CVn phenomenon are expected to be dark in the far-UV (FUV) and bright in the optical wavebands \citep{Mikulasek2019}, resulting in the vEHBs being fainter in FUV, depending on the spot properties and showing higher variability in the FUV. Understanding the FUV properties of EHB stars as a function of their rotation will help in throwing more light on our emerging understanding of EHB stars. Moreover, observing such hot stars in the UV is advantageous because the contribution from cooler stellar populations such as main sequence (MS) and RGB stars gets suppressed, thereby considerably reducing the stellar crowding in the central regions of GCs, where many of the EHB stars are located.

In this study, we present the far-UV and near-UV photometric analysis and spectral energy distributions (SEDs) of 12 vEHB stars and 3 vEHB candidates newly discovered in the GC NGC\,2808 by \citet{Momany2020}. This GC has an age = $10.9 \pm 0.7$ Gyrs \citep{Massari2016}, [Fe/H] = $-$1.14 dex and is located at a distance of 9.6 kpc \citep[2010 edition, H96]{Harris1996}. We make use of the archival UV data from the \textit{AstroSat}/UVIT along with UV-optical data from the {\it Hubble Space Telescope} UV Globular Cluster Survey (HUGS) catalogue \citep{ Nardiello2018,Piotto2015}, optical data from ground-based telescopes \citep{Stetson2019} and others available in literature. 
\citet{Jain2019} used the UVIT data for this cluster to study the photometric gaps in the colour-magnitude diagrams (CMDs) and the multiple stellar populations. Prabhu {\em et al.} (2020, submitted) presented a detailed analysis of the stars in the pHB phase in this cluster utilizing the UVIT data in combination with other multiwavelength datasets. In this work, we study the UV properties of the spotted EHB stars in the cluster.

The paper is laid out as follows. Section \ref{obs} gives the details of the observations and data reduction procedure. The UV photometric analysis of vEHB stars is described in Section \ref{vehbs} The SEDs of these stars and the derived results are presented in Section \ref{seds} The results are discussed in Section \ref{discussion} and a summary is presented in Section \ref{summary}

\section{Observations and Data Reduction} \label{obs}
In this study, we utilized the archival \textit{AstroSat}/UVIT data for NGC\,2808. The UVIT is made up of two 38 cm diameter telescopes, one for FUV ($\lambda = 1300-1800$ \AA) band pass, and the other for NUV ($\lambda = 2000-3000$ \AA) and VIS ($\lambda = 3200-5500$ \AA) band passes. The data from the VIS channel are used mainly for the drift-correction of images. The field of view (FOV) of UVIT is circular with a diameter of $28'$. For more details regarding the instrument and calibration, refer \citet{Tandon2017}.

We used the data in two FUV (F154W and F169M) and four NUV filters (N242W, N245M, N263M and N279N). The images in these filters were generated using the CCDLAB software package \citep{Postma2017} by correcting for the spacecraft drift, geometrical distortions and flat-field illumination. The final science-ready images were created by aligning and merging the images corresponding to different orbits.    

We then executed crowded field photometry on these images using the DAOPHOT software package of IRAF/NOAO \citep{Stetson1987}. The detailed steps of the procedure can be found in Prabhu {\em et al.} (2020, submitted). The final magnitudes (AB system) in all the filters were corrected for extinction by choosing the reddening value of $E(B-V)$ = 0.22 mag \citep[2010 edition]{Harris1996} and the ratio of total to selective extinction, $R_V$ = 3.1. To calculate the extinction coefficients in various filters, the reddening law of \citet{Cardelli1989} was employed.

For identifying the different classes of stars detected in the UVIT images and to obtain their cluster membership probabilities, the stars within the central $\sim$ $2.7'$ $\times$ $2.7'$ region (inner region) were cross-matched with the HUGS catalogue as described in Prabhu {\em et al.} (2020, submitted). The HUGS catalogue consists of data in five filters, namely, the WFC3/UVIS F275W (NUV), F336W ({\it U}) and F438W ({\it B}) filters, along with the ACS/WFC F606W ({\it V}) and F814W ({\it I}) filters. 

The UVIT covers a much larger FOV as compared to the  {\it Hubble Space Telescope} ({\it HST}). Thus, in order to identify stars in the outer region (region outside HUGS coverage) and to select the cluster members among them, the UVIT data were combined with other datasets. The membership probabilities of the stars were estimated based on the $Gaia$ DR2 proper-motion data using the technique of \citet{Singh2020}. The Johnson-Cousins $UBVRI$ photometry of cluster members thus selected, were obtained by a cross-match with the data from \citet{Stetson2019}. In order to plot the stars in the inner and outer regions in the same colour and magnitude plane, we transformed the Johnson $V$ magnitudes into the equivalent $HST$ ACS/WFC filter using the transformation equations of \citet{Sirianni2005}.

\section{Spotted EHB stars} \label{vehbs}

\begin{figure} [!htb]

\includegraphics[width=\columnwidth]{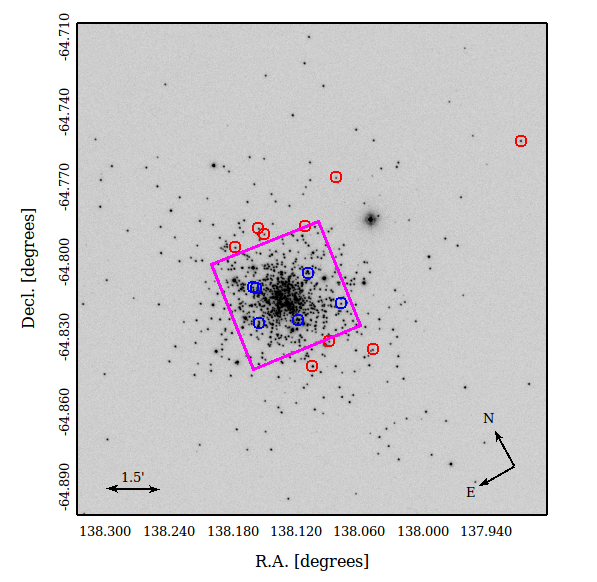}
     
\caption{The vEHB stars marked over the UVIT F154W image of NGC\,2808. The magenta region marks the central $\sim$ $2.7'$ $\times$ $2.7'$ region (inner region) covered by {\it HST} and the stars within this region are encircled in blue. The stars lying in the outer region are encircled in red.} 
\label{spatial_vEHBs}
\end{figure}

The 12 vEHB stars and the 3 vEHB candidates (vEHB-C) from \citet{Momany2020} were cross-identified in our UV catalogue. We note that some of these vEHB stars did not meet the membership-probability cut-off. Nevertheless, we include them in our study, as they are considered by \citet{Momany2020} and this is a follow up study of their objects. Figure\,\ref{spatial_vEHBs} shows the spatial locations of these vEHBs (including candidates) marked over the UVIT F154W image. The stars within the inner region, covered by the $HST$ WFC3/UVIS, are encircled in blue and the other stars in red. We would also like to point out that only one of the vEHBs located in the inner region (vEHB-7) is affected by neighbour contamination in the UVIT images. Hence, we do not use the UVIT data for further analysis of this star.

\subsection{vEHBs in FUV-optical CMD}

\begin{figure*}[!htb]
\makebox[\linewidth]
{
\includegraphics[width=1.1\textwidth]{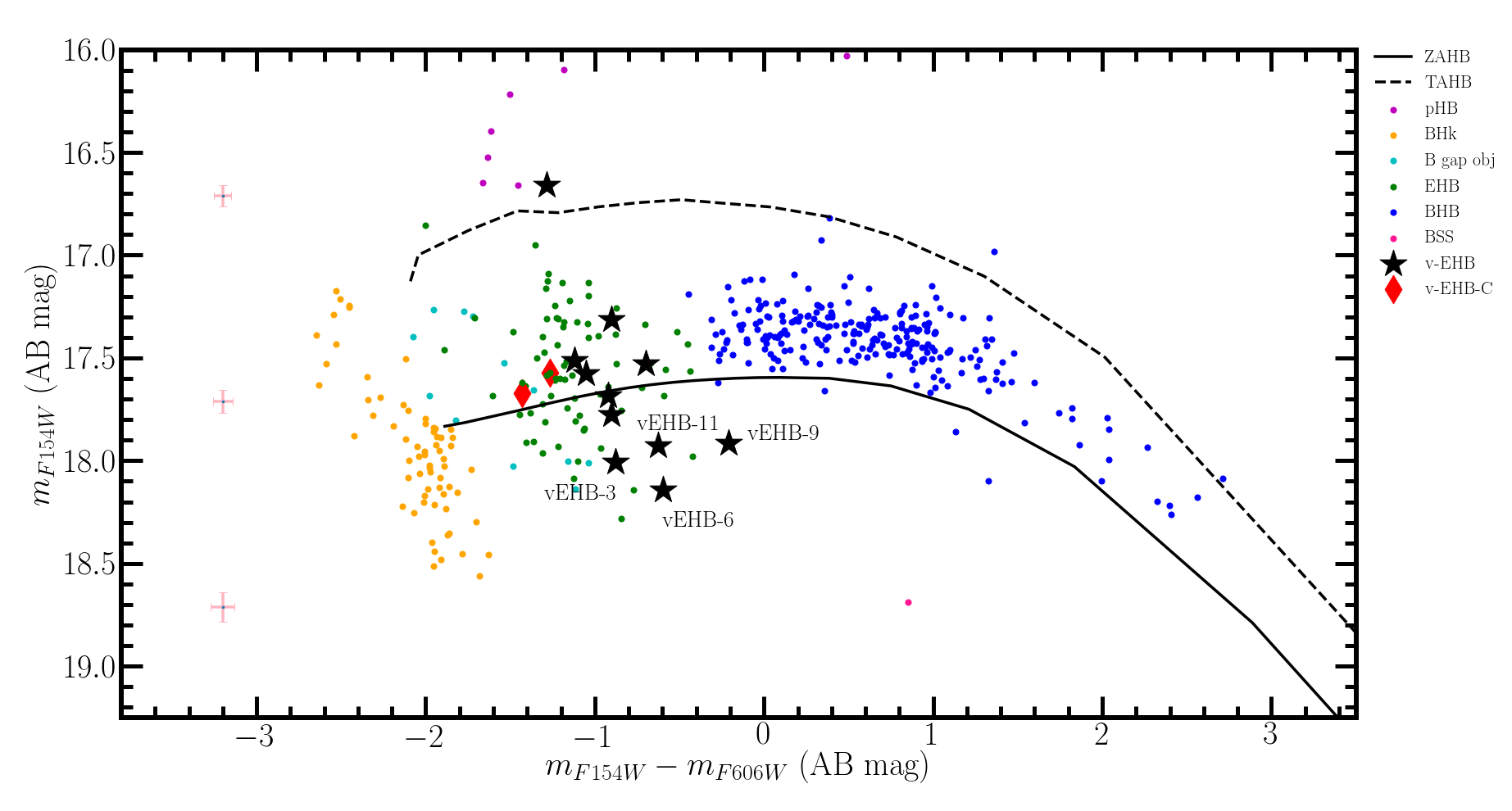}
}
\caption{The $m_{F154W} - m_{F606W}$ vs $m_{F154W}$ CMD for the cluster member stars common in F154W UVIT filter and other catalogues (HUGS and ground-based optical data). The black solid and dashed lines represents the BaSTI ZAHB and TAHB models with [Fe/H] = $-$0.9 dex, respectively. The photometric errors in magnitude and color are also shown along the left side of the plot. The 4 faintest vEHB stars are annotated.}
\label{baf2_V_cmd}
\end{figure*}

Figure\,\ref{baf2_V_cmd} shows the $m_{F154W} - m_{F606W}$ vs $m_{F154W}$ CMD for the cross-matched stars in our catalogue. To create this CMD, the F606W magnitudes are converted from the VEGA magnitude system to the AB system using appropriate conversion factors.\footnote{ \url{http://waps.cfa.harvard.edu/MIST/BC_tables/zeropoints.txt}} The HB sequence of the CMDs are overlaid with the updated BaSTI (a Bag of Stellar Tracks and Isochrones) theoretical zero-age HB (ZAHB) and terminal-age HB (TAHB; end of He burning phase) models from \citet{Hidalgo2018}. These models correspond to metallicity [Fe/H] = $-$0.9 dex, He abundance = 0.249, solar scaled [$\alpha$/Fe] = 0.0,  and no convective overshoot. The models also account for the effects due to atomic diffusion. The different classes of HB stars (BHk, EHB, BHB), B gap objects, the pHB stars and the BSSs are indicated with distinct colours. 

In the figure, the vEHB stars are shown with black star symbols and the candidates with red diamonds. We do not include vEHB-7 in this plot for the reason mentioned earlier. One vEHB (vEHB-10) is seen to be located just above the TAHB sequence. This star along with vEHB-7 have been previously identified as pHB candidates by Prabhu {\em et al.} (2020, submitted) with vEHB-7 corresponding to Star 21 and vEHB-10 corresponding to Star 33 in their catalogue. One of these stars (vEHB-10) is also listed as an AGB-manqu\'e star (NGC\,2808-594) in the study by \citet{Schiavon2012}. 

In the CMD, several EHB stars are observed to be sub-luminous with respect to the ZAHB. About half of the vEHB stars belong to this sub-luminous EHB population. 

\subsection{Correlation between UV magnitudes and periods of vEHBs}

\begin{figure*}[!htbp]
\makebox[\linewidth]
{
\includegraphics[width=1.1\textwidth]{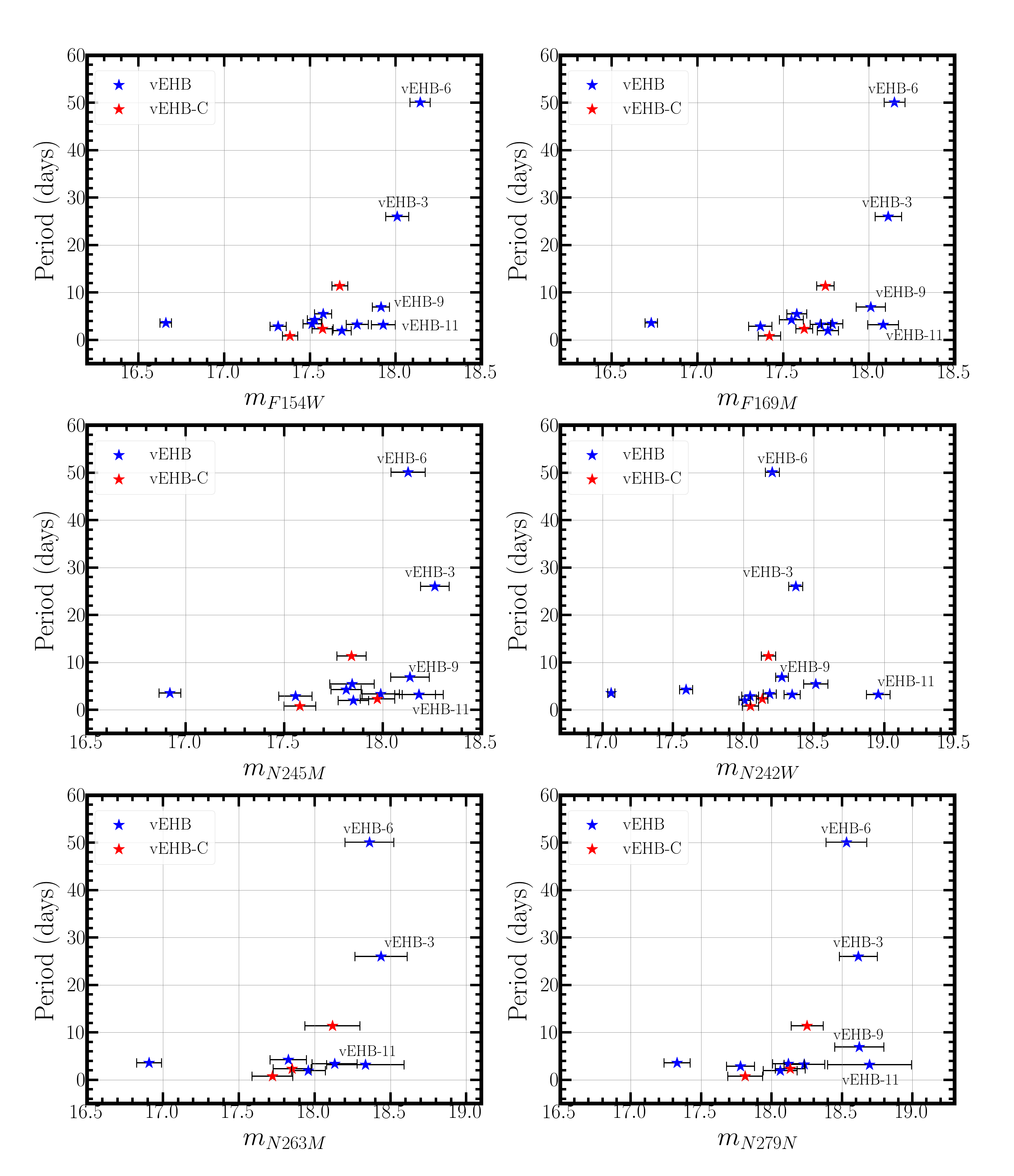}
}
\caption{The magnitudes (extinction corrected) of vEHB stars along with their photometric errors in different UVIT filters, plotted against their periods from \citet{Momany2020}. In the filters N263M and N279N, not all vEHBs are detected. The star vEHB-7 is excluded because it is affected by neighbour contamination. The IDs of the four stars with the faintest UV magnitudes are marked.}
\label{pd_mag}
\end{figure*}

To check whether any correlation exists between the UV magnitudes and periods of these vEHBs, we plotted them against each other in Figure\,\ref{pd_mag}.  The periods of these stars are obtained from \citet{Momany2020}. We note the following from the figure: in general, stars with the longest periods are fainter in the FUV and NUV. There is a tentative correlation between period and FUV magnitudes (F154W and F169M), and also within errors in the NUV magnitudes (N242W, N245M, N263M and N279N) such that longer period stars are fainter in the UV, except for two stars, which deviate from this trend. That is, among the 4 stars that are UV faint and found below the ZAHB (vEHB-3, vEHB-6, vEHB-9 and vEHB-11), vEHB-3 and vEHB-6 have longer periods, whereas the other two have short periods. This is noticed in all the sub-plots shown in Figure\,\ref{pd_mag}. Therefore, from the figures, even though we notice a tentative correlation, we are unable to make a definite correlation. 

In order to further understand this aspect and to explore the correlations (if any) between the fundamental parameters of vEHBs and their periods, we generated their SEDs as explained in the next section. 

\section{Spectral Energy Distributions} \label{seds}

We exploited the available multiwavelength photometric data for the vEHB stars to construct their SEDs and derive parameters such as ${\it T_{eff}}$, luminosity ($L$), and radius ($R$). We used a virtual observatory tool, VOSA (VO SED Analyser; \citealt{Bayo2008}), for this purpose. VOSA produces synthetic photometric points for the selected theoretical models using the response curves of the required photometric filters. In order to estimate the best-fit parameters for the SEDs, the observed and synthetic photometric points are compared using the $\chi^{2}$ minimization technique. The $\chi_{red}^{2}$ value is found using the relation, 

\begin{center}
\begin{equation}
\chi_{red}^{2} = \displaystyle \frac{1}{N-N_{f}} \sum_{i=1}^{N} \Bigg\{ \frac{(F_{o,i}-M_{d}F_{m,i})^{2}}{\sigma_{o,i}^{2}}\Bigg\}
\end{equation}
\end{center}

where {\it N} is the number of photometric points, ${\it N_{f}}$ is the number of fitted parameters for the model, ${\it F_{o,i}}$ is the observed flux, ${\it F_{m,i}}$ is the flux predicted by the theoretical model, ${\it M_{d}} = {\it (\frac{R}{D})^{2}}$ is the multiplicative dilution factor (where $R$ is the radius of the star and {\it D} is the distance to the star) and ${\it \sigma_{o,i}}$ is the error in the observed flux. For all the vEHB stars, we assumed {\it D} = 9.6 kpc and $E(B-V)$ = 0.22 mag \citep[2010 edition]{Harris1996}. For accounting the extinction in the observed photometric data points, VOSA uses the Fitzpatrick reddening relation \citep{Fitzpatrick1999}. 

\begin{figure*}[!htbp]
\makebox[\textwidth]
{
\subfloat{\includegraphics[width=0.55\textwidth]{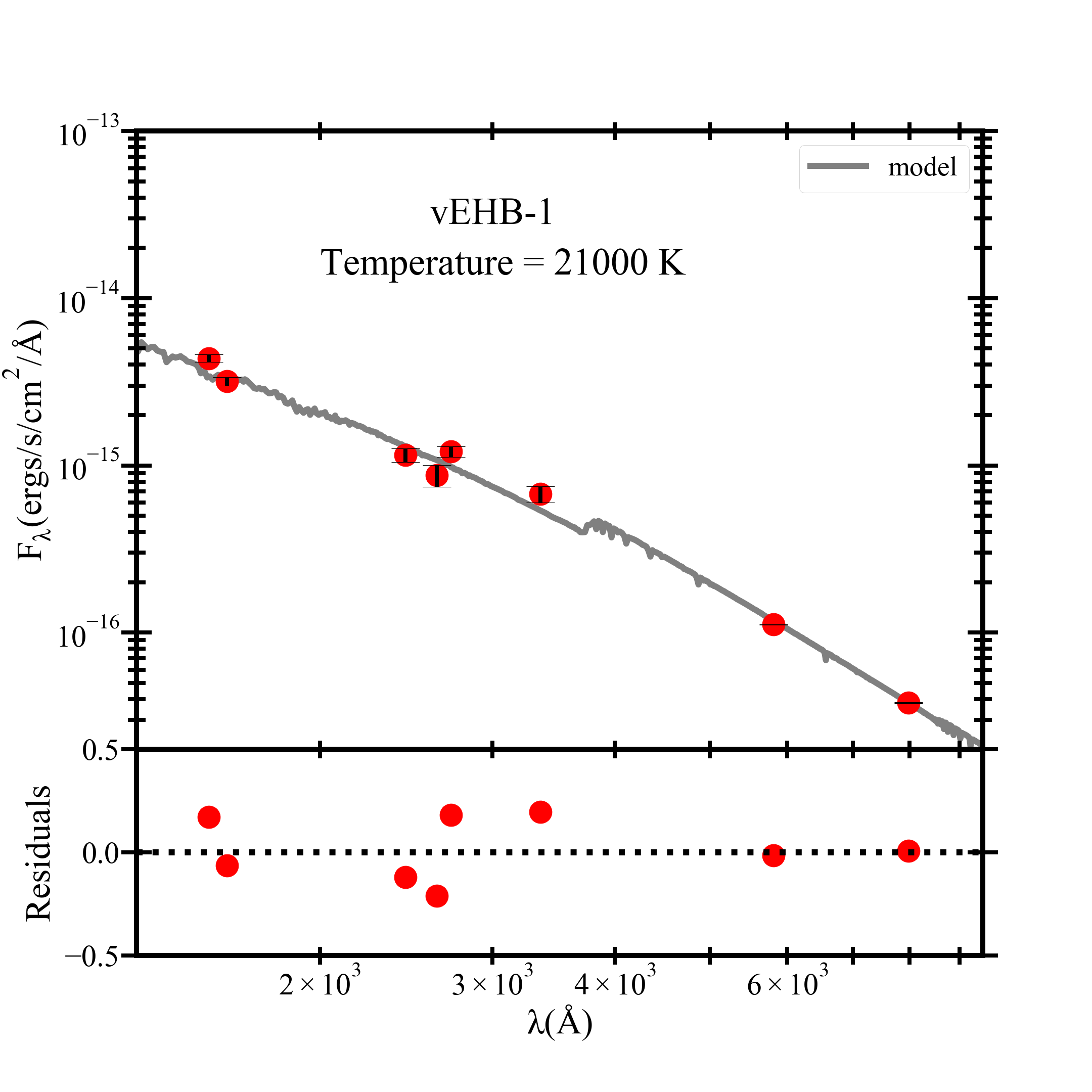}}
\subfloat{\includegraphics[width=0.55\textwidth]{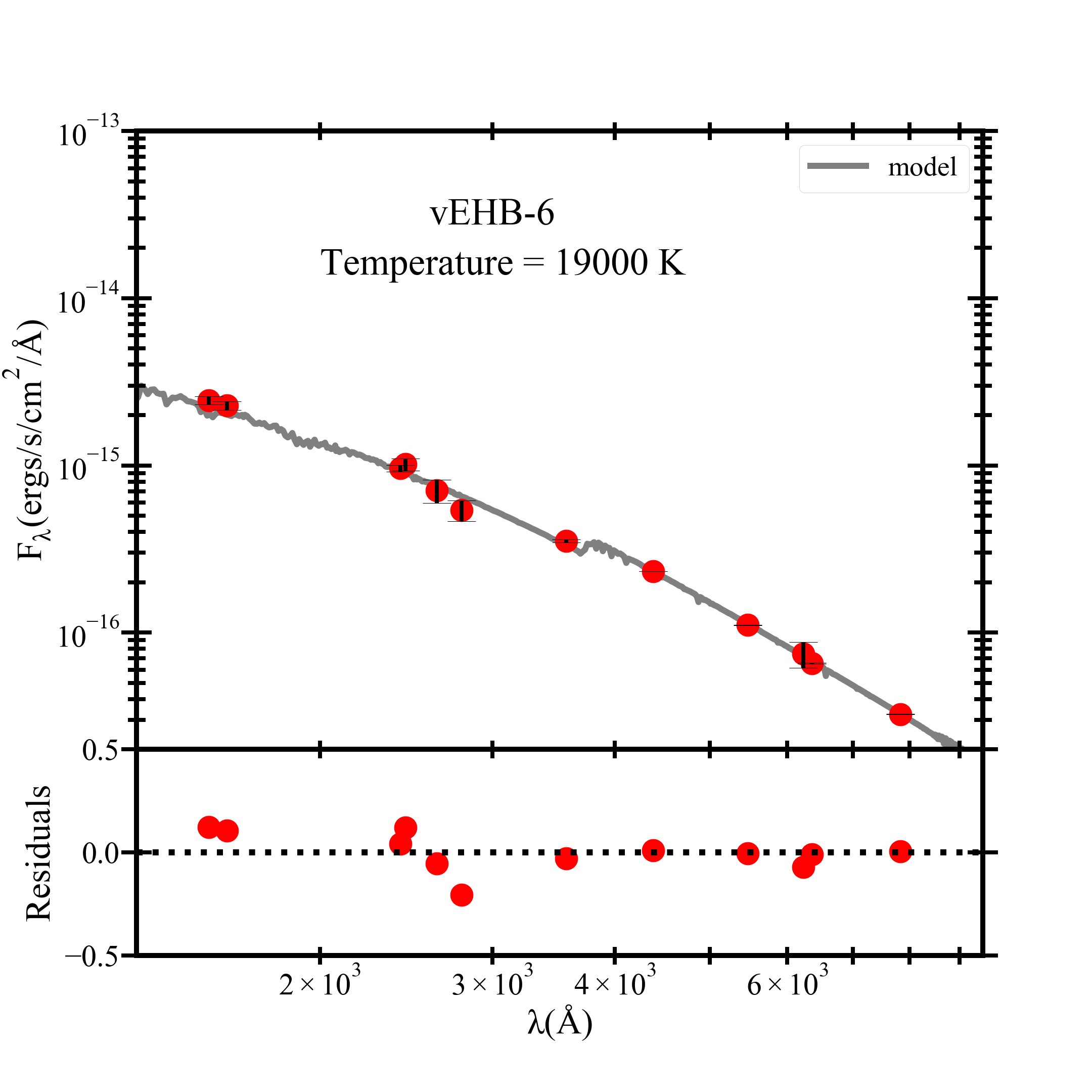}}
}
\caption{The SEDs for two of the vEHBs, namely vEHB-1 and vEHB-6, after extinction correction. The gray line shows the model spectrum. The residuals of SED fit are shown in the bottom panels of both plots.}
\label{SED_eg}
\end{figure*}

For the vEHB stars located within the central $\sim$ $2.7'$ $\times$ $2.7'$ region of the cluster, we used the photometric data from UVIT and HUGS catalogues to generate the SEDs. In the case of stars located outside this region, we used the UVIT data in combination with $UBVRI$ photometry from \citet{Stetson2019}, and other available VO photometric data from VOSA. The Kurucz stellar atmospheric models \citep{Castelli1997, Castelli2003} were used to fit the SEDs. The free parameters in these models are : log $\textit{g}$ with a range 0.0 to 5.0, [Fe/H] with a range $-$2.5 to 0.5 and ${\it T_{eff}}$ ranging from 3500 to 50000 K.  In order to fit the SEDs with these models, the value of [Fe/H] was fixed at $-$1.0 dex, close to the value for the cluster.  The best-fit parameters derived from the SED analysis for these stars are tabulated in Table \ref{tab:sed_params}. The parameter errors quoted in the table were estimated as half the grid step, around the best-fit value. We refrain from tabulating the best-fit log $\textit{g}$ values as SED analysis does not give accurate values of this parameter. Figure\,\ref{SED_eg} shows examples of SED fits for vEHB-1 and vEHB-6 with fit residuals. The residuals were calculated for each photometric point as follows :

\begin{center}
\begin{equation}
Residual = \frac{F_{o} - F_{m}}{F_{o}}
\end{equation}
\end{center}

where $F_{o}$ and $F_{m}$ are the observed and the theoretical model fluxes corresponding to the photometric points.


\begin{table*}[!htbp]
\caption{Table with the results of SED fitting procedure for the 15 vEHBs in the cluster. Columns 1 and 2 show the star IDs and periods of the vEHBs from \citet{Momany2020}.  The estimated parameters such as effective temperature, luminosity (in solar units) and radius (in solar units) of the stars and the errors in these parameters are tabulated in Columns 3 to 5. The errors are obtained as half the grid step, around the best-fit value. Columns 6 and 7, respectively, show the reduced chi square value corresponding to the fit and the number of photometric points used for fitting. Column 8 shows the spatial location (S.L) of the stars where `inner' stands for the star lying within the HST FOV and `outer' stands for star lying outside this region. The membership probability (M.P) of each star is indicated in the last column.}
\label{tab:sed_params}
\resizebox{\textwidth}{!}{%
\begin{tabular}{lcccccccl}
\topline
{ID} & Period$^{(a)}$ & ${\it T_{eff}}$ & ${\it \frac{L}{L_{\odot}}}$ & ${\it \frac{R}{R_{\odot}}}$ & $\chi_{red}^{2}$ & ${\it N_{fit}}$ & S.L. & M.P. \\ \hline
 &    (days)    & (K)             &                  &                 &      &    &       & {($\%$)} \\ \hline
vEHB-1               & 3.38683880 & 21000 $\pm$ 500 & 20.56 $\pm$ 0.19 & 0.34 $\pm$ 0.02 & 7.05 & 8  & inner & 97.3                       \\
vEHB-2               & 5.47704961 & 22000 $\pm$ 500 & 22.11$\pm$ 0.11  & 0.32 $\pm$ 0.01 & 2.09 & 8  & inner & 97.3                       \\
vEHB-3               & 26.01823489 & 21000 $\pm$ 500 & 13.44 $\pm$ 0.21 & 0.28 $\pm$ 0.01 & 4.09 & 12 & outer & 0                          \\
vEHB-4               & 1.97628738 & 23000 $\pm$ 500 & 22.15 $\pm$ 0.22 & 0.30 $\pm$ 0.01 & 9.57 & 10 & outer & 0                          \\
vEHB-5               & 3.23990003 & 22000 $\pm$ 500 & 18.92 $\pm$ 0.11 & 0.30 $\pm$ 0.01 & 5.10 & 8  & outer & ---                        \\
vEHB-6               & 50.10395158 & 19000 $\pm$ 500 & 12.54 $\pm$ 0.21 & 0.32 $\pm$ 0.02 & 4.20 & 12 & outer & 98.9                       \\
vEHB-7*               & 3.02583807 & 24000 $\pm$ 500 & 49.30 $\pm$ 0.08 & 0.41 $\pm$ 0.02 & 3.75 & 5  & inner & 96.6                       \\
vEHB-8               & 2.89086189 & 21000 $\pm$ 500 & 28.77 $\pm$ 0.17 & 0.41 $\pm$ 0.02 & 3.06 & 9  & inner & 98                         \\
vEHB-9               & 6.90595098 & 18000 $\pm$ 500 & 14.96 $\pm$ 0.25 & 0.40 $\pm$ 0.02 & 2.57 & 8  & outer & 0                          \\
vEHB-10*              & 3.58179363 & 26000 $\pm$ 500 & 53.26 $\pm$ 0.51 & 0.36 $\pm$ 0.01 & 2.39 & 11 & outer & 92.6                       \\
vEHB-11              & 3.19599005 & 17000 $\pm$ 500 & 16.35 $\pm$ 0.16 & 0.47 $\pm$ 0.03 & 1.75 & 7  & inner & 59.4                       \\ 
vEHB-12              & 4.26386105 & 21000 $\pm$ 500 & 25.24 $\pm$ 0.48 & 0.38 $\pm$ 0.02 & 8.99 & 15 & outer & 99.3                       \\ \hline
\multicolumn{9}{c}{Candidate vEHBs}
   \\ \hline
vEHB-13              & 0.80305948 & 21000 $\pm$ 500 & 27.57 $\pm$ 0.17 & 0.40 $\pm$ 0.02 & 8.53 & 9  & inner & 97.9                       \\
vEHB-14              & 11.37292900 & 24000 $\pm$ 500 & 18.75 $\pm$ 0.22 & 0.25 $\pm$ 0.01 & 4.00 & 11 & outer & 0                          \\
vEHB-15              & 2.32221267 & 25000 $\pm$ 500 & 21.94 $\pm$ 0.16 & 0.25 $\pm$ 0.01 & 3.00 & 8  & outer & 96.1                       \\ \hline
\end{tabular}%
}
\tablenotes{$^{(a)}$ From \citet{Momany2020}. $^{(*)}$ These stars are classified as pHB stars by Prabhu {\em et al} (2020, submitted). vEHB-10 is categorized as a pHB star also by \citet{Schiavon2012}.}
\end{table*}


\begin{figure*}[!htbp]
\makebox[\textwidth]
{
\subfloat{\includegraphics[width=0.55\textwidth]{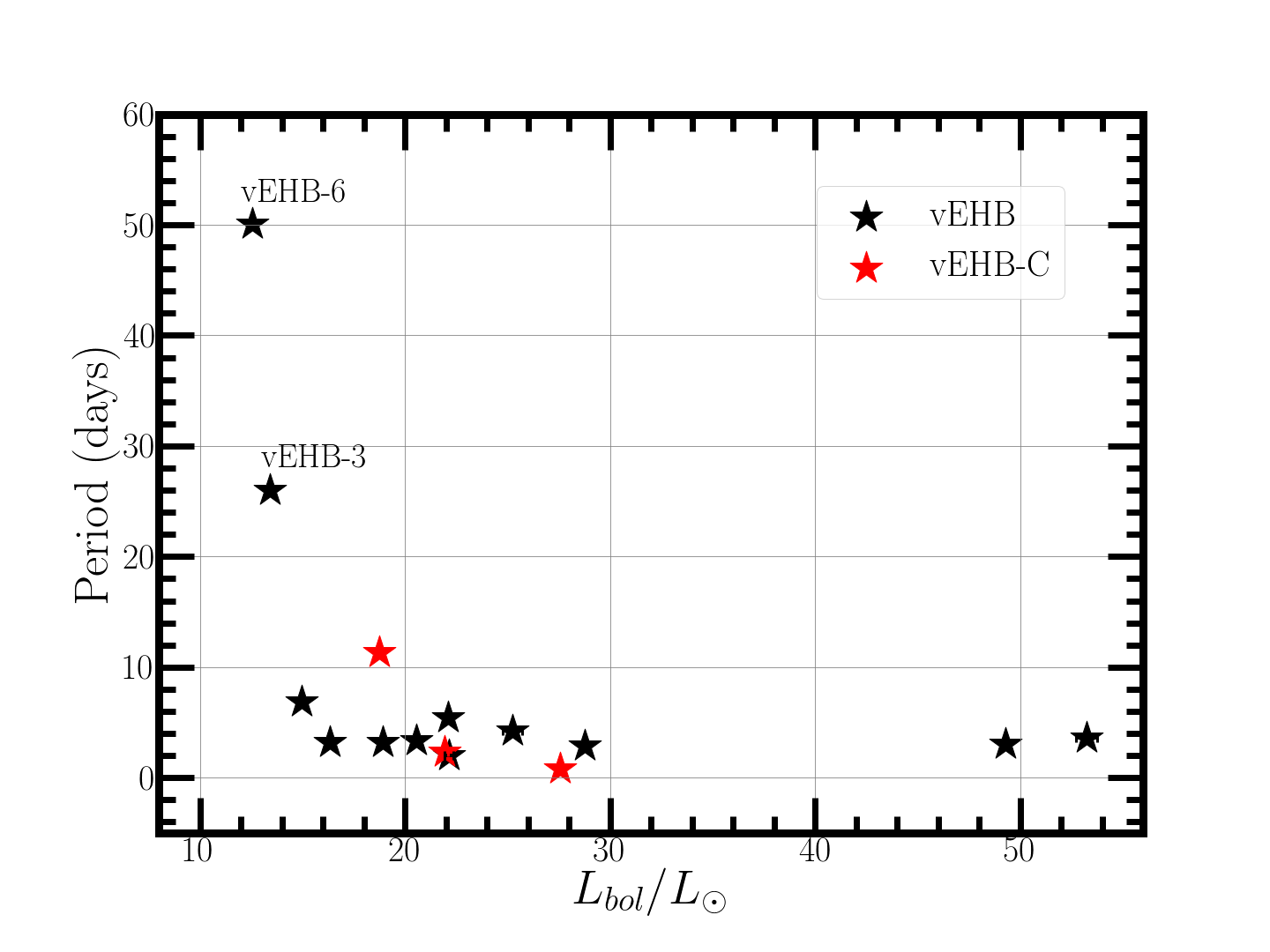}}
\subfloat{\includegraphics[width=0.55\textwidth]{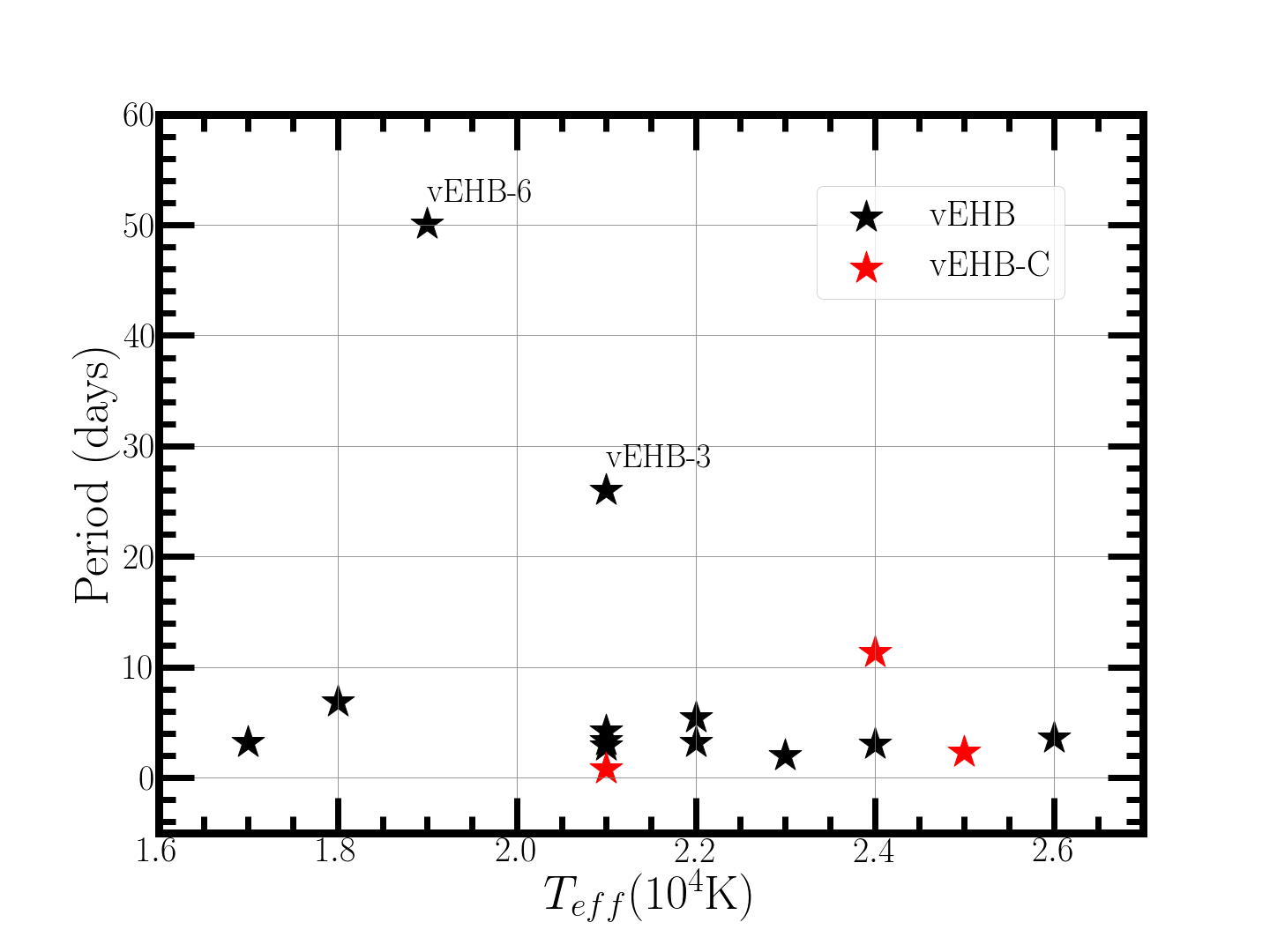}}
}
\caption{The $L/L_{\odot}$ and $T_{eff}$ derived from SED fitting technique, plotted against the periods of all the vEHB stars. The longest period variables are marked in the plots.}
\label{lum_vs_pd}
\end{figure*}

We now plot the derived fundamental parameters, $L$ and $T_{eff}$, against the periods of the vEHB stars in Figure\,\ref{lum_vs_pd}. From the right panel, we hardly see any correlation between the  $T_{eff}$ and period. The left panel shows the same tentative trend that was observed earlier, that the vEHBs with the longest periods have the smallest bolometric luminosities. The two stars which slightly deviate from the trend are also the same.  Therefore, we conclude there is a tentative correlation between the luminosity and period among the EHB stars.


\section{Discussion} \label{discussion}
The UV follow-up study of the newly discovered spotted EHB stars in the GC NGC\,2808 is presented here. We utilize the UVIT data complemented by the data from HUGS catalogue for the inner region of the cluster and the ground-based optical photometry catalogue of \citet{Stetson2019} for the outer region. From the $m_{F154W} - m_{F606W}$ vs $m_{F154W}$ CMD, we find that about half of the vEHB stars belong to the EHB population which is sub-luminous with reference to the theoretical ZAHB model. The faintest among these stars (vEHB-6 and vEHB-3) are those with the longest periods. However, the location of these stars in the CMD may not be an effect of their rotation as we detect two other stars with shorter periods which have similar F154W magnitudes within errors. 

The SED analysis of vEHB stars leads to an inference that the slowest rotators also have the smallest bolometric luminosities. We also find that two of the vEHBs have very high luminosities (vEHB-7 and vEHB-10). These stars have been classified as AGB-manqu\'e stars in the works of \citet{Schiavon2012} and Prabhu {\em et al} (2020, submitted). Also, since the amplitude of variability of all the vEHB stars is very small ($\Delta U_{Johnson} \sim 0.04-0.22$ mag), it highly improbable that the results obtained in this study are significantly affected due to the variability. 

\citet{RecioBlanco2002} obtained the rotational velocities of hot HB stars in this GC. Their analysis showed that about 20$\%$ of the HB stars hotter than $T_{eff}$ $\sim$ 11,500 K are slow rotators ($v$ sin $i$ $<$ 2 km s$^{-1}$). We plan to obtain the catalogue of these slow rotating hot HB stars from the authors and locate them in our UV study. This might also give some clues regarding these stars and their UV properties.

In future, we also plan to look into the NUV and FUV variability of these spotted stars using the UVIT data by generating light curves.

\section{Summary} \label{summary}
1. We present the \textit{AstroSat}/UVIT photometric analysis of the recently discovered vEHB stars in the GC NGC\,2808. The UVIT data is used in combination with the data from HUGS, and ground-based optical photometry catalogues. The UV study of these objects is crucial because the $\alpha^{2}$\,CVn mechanism which is responsible for the detected variability in EHB stars results in fainter UV magnitude depending on spot properties and higher variability in the UV wavebands than in the optical. 

2. We present the FUV-optical CMD of the UV-bright cluster members along with all the reported vEHB stars. The location of the HB sequence is compared with the theoretical ZAHB and TAHB models. We find that about half of the vEHB stars belong to the EHB population that is sub-luminous in comparison with the theoretical ZAHB. 

3. A plot of the UV magnitudes versus the periods of the vEHB stars shows that the two longest period variables (vEHB-6 and vEHB-3) are the faintest in the two FUV wavebands and within errors in the four NUV 
wavebands.

4. SED fitting technique was adopted to estimate parameters such as ${\it T_{eff}}$, $R/R_{\odot}$ and $L/L_{\odot}$ of these stars. The plot of $L/L_{\odot}$ against the periods shows that the two longest period vEHBs also have the smallest bolometric luminosities among the sample. 

5. A detailed study of the UV variability of these spotted stars using UVIT data will be carried out in the near future.




\section*{Acknowledgements}
We gratefully acknowledge Y. Momany for the useful discussions. This publication utilizes the data from {\it AstroSat} mission's UVIT, which is archived at the Indian Space Science Data Centre (ISSDC). The UVIT project is a result of collaboration between IIA, Bengaluru, IUCAA, Pune, TIFR, Mumbai, several centers of ISRO, and CSA. This research made use of VOSA, developed under
the Spanish Virtual Observatory project supported by the
Spanish MINECO through grant AyA2017-84089. This research also made use of Topcat \citep{Taylor2005}, ``Aladin sky atlas" developed at CDS, Strasbourg Observatory, France \citep{Bonnarel2000,Boch2014}, Matplotlib, NumPy, SciPy and pandas. 

\vspace{-1em}

\bibliography{ngc2808_EHB}{}

\end{document}